\documentclass[12pt]{article}
\usepackage[left=0.85in,right=0.85in,top=1in,bottom=1in]{geometry}
\usepackage{setspace}
\usepackage{etoolbox}
\AtBeginEnvironment{thebibliography}{\singlespacing}
\usepackage{amsmath,amsthm,amssymb}
\usepackage{microtype}
\usepackage[linesnumbered,ruled]{algorithm2e}
\SetKwInOut{Parameter}{Parameters}
\usepackage{float}
\usepackage{bm}
\usepackage{graphicx}
\usepackage{enumerate}
\usepackage{hyperref}
\usepackage[numbers]{natbib} 
\usepackage{url} % not crucial - just used below for the URL
\usepackage[table]{xcolor}
\usepackage{subcaption}
\usepackage{caption}
\newcommand{\Title}{Causality-Encoded Diffusion Models for Interventional Sampling and Edge Inference}

\hypersetup{
    hidelinks,
    pdftitle={\Title},
    pdfauthor={Li Chen, Xiaotong Shen, Wei Pan}
}
\usepackage{orcidlink}
\newcommand{\blind}{1}
\usepackage{booktabs}
\usepackage{tikz}
\usetikzlibrary{arrows.meta,positioning}

\newtheorem{theorem}{Theorem}
\newtheorem{lemma}{Lemma}

\newtheorem{proposition}{Proposition}

\theoremstyle{definition}

\newtheorem{definition}{Definition}

\newtheorem{assumption}{Assumption}

\newcommand{\comment}[1]{}

\newcommand{\E}{\operatorname{\mathbb{E}}} % expectation
\newcommand{\pa}{{\mathrm{pa}}}
\newcommand{\doop}{\mathrm{do}}
\begin{document}

%%%%%%%%%%%%%%%%%%%%%%%%%%%%%%%%%%%%%%%%%%%%%%%%%%%%%%%%%%%%%%%%%%%%%%%%%%%%%%

\if1\blind
{
  \title{\bf \Title}
    \author{Li Chen\thanks{School of Statistics, University of Minnesota, Minneapolis, MN 55455.}
    \and 
    Xiaotong Shen\textsuperscript{\orcidlink{0000-0003-1300-1451}}\thanks{To whom correspondence should be addressed.
    School of Statistics, University of Minnesota, Minneapolis, MN 55455. Email: \url{xshen@umn.edu}.}
    \and 
    Wei Pan\thanks{Division of Biostatistics, University of Minnesota, Minneapolis, MN 55455.}}
    \date{}
  \maketitle
} \fi

\if0\blind
{
  \bigskip
  \bigskip
  \bigskip
  \title{\bf \Title}
  \date{}
  \maketitle
  \medskip
} \fi

\bigskip

\begin{abstract}
Standard diffusion models are flexible estimators of complex distributions, but they do not encode causal structures and therefore do not by themselves support causal analysis. We propose a causality-encoded diffusion framework that incorporates a known directed acyclic graph by training conditional diffusion models consistent with the graph factorisation. The resulting sampler approximately recovers the observational distribution and enables interventional sampling by fixing intervened variables while propagating effects through the graph during reverse diffusion. Building on this interventional simulator, we develop a resampling-based test for directed edges that generates null replicates under a candidate graph. We establish convergence guarantees for observational and interventional distribution estimation, with rates governed by the maximum local dimension rather than the ambient dimension, and prove asymptotic control of type~I error for the edge test. Simulations show improved interventional distribution recovery relative to baselines, with near-nominal size and favourable power in inference. An application to flow cytometry data demonstrates practical utility of the proposed method in assessing disputed signalling linkages.
\end{abstract}
\noindent%

\section{Introduction}

Diffusion models \citep{sohl2015deep,ho2020denoising,song2021scorebased} have emerged as a powerful class of generative models, achieving state-of-the-art performance across a wide range of applications, including imaging \citep{ho2020denoising} and scientific-data synthesis \citep{shang2025predicting}. From a statistical perspective, they can be viewed as flexible nonparametric estimators of a (conditional) distribution via score estimation and reverse-time stochastic differential equations (SDEs) \citep{anderson1982reverse,kumar2025likelihood}. Despite this expressive power, standard diffusion models are typically \emph{causality-agnostic}: they learn a joint law without encoding the directional asymmetries required for causal interpretation. As a consequence, they do not, on their own, provide principled answers to interventional queries or support broader causal analyses, which are central to structural causal models (SCMs) \citep{pearl2009causality}.

When a causal ordering (or a directed acyclic graph) is available, it is natural to construct generative procedures that sample variables sequentially according to the causal factorisation. Such iterative, ordering-respecting approaches have been proposed using a variety of generative models, including generative adversarial networks \citep{kocaoglu2018causalgan}, variational autoencoders \citep{karimi2020algorithmic}, normalising flows \citep{pawlowski2020deep}, and diffusion-based constructions such as DDIM \citep{chao2023modeling}. However, a rigorous statistical understanding of the advantages of exploiting such causal structure and the inferential use of the resulting generator remain less developed.

In parallel, a recent line of work has sought to alleviate the curse of dimensionality in diffusion models. For instance, \citet{huang2024denoising,li2025d,potaptchik2025linear} provide convergence guarantees suggesting that sampling complexity may scale with an intrinsic, rather than ambient, dimension under suitable conditions. These results primarily address sampling complexity, whilst accurate score estimation can remain a bottleneck. Complementarily, \citet{chen2023score,tang2024adaptivity} study score approximation and distribution recovery under manifold structure, deriving rates that reflect the underlying geometry, though their manifold assumptions may be restrictive in practice. \citet{tian2024enhancing} instead considers dimension reduction through transfer learning, assuming a shared low-dimensional latent representation between target and auxiliary data. However, identifying auxiliary sources that enable positive transfer can be challenging. Our work offers a different perspective: when a causal structure is available, we show theoretically that distribution recovery for an iterative, DAG-respecting generative procedure is governed by a \emph{maximum local dimension} (a node together with its parents), rather than by the ambient dimension. This yields a principled route to dimension reduction that is structural rather than geometric.

A separate but related challenge concerns statistical inference for directed edges in DAG models. Recent work has developed methods that attach explicit measures of uncertainty to individual edges, including debiased estimators and confidence intervals for edge weights \citep{jankova2018inference}, constrained likelihood ratio tests \citep{li2020likelihood,chen2024discovery}, and neural-network-based hypothesis testing \citep{shi2024testing}. These approaches often rely on Gaussian linear or additive-noise assumptions, and extending them to general nonlinear and non-additive settings is non-trivial. Another broad literature studies conditional independence testing (CIT); see, for example, \citet{candes2018panning,berrett2020conditional,tansey2022hrt}. Leveraging diffusion models as powerful conditional generative tools, we use resampling-based CIT as a core building block for edge testing. Combined with a new multivariate conditional dependence coefficient tailored to our null-generation mechanism, our inference procedure imposes minimal restrictions on the underlying DAG model, substantially broadening the scope of existing DAG inference methods.

The contributions of this paper are threefold:
\begin{enumerate}[(a)]
    \item \textbf{Methodology.} We introduce \emph{Causality-Encoded Diffusion Models} (CEDM), a score-based generative framework that respects causal structure, together with samplers for both observational and $\doop$-interventional distributions. Building on this $\doop$-sampling capability, we develop \emph{CEDM-based inference} (CEDMI), a resampling-based procedure that fits a CEDM under the relevant null graph and uses the resulting generator to test targeted directed edges.
    \item \textbf{Statistical guarantees.} We establish convergence guarantees for observational and interventional sampling error in total variation distance, with rates governed by the \emph{maximum local dimension} rather than the ambient dimension. We also prove asymptotic type~I error control for CEDMI.
    \item \textbf{Empirical validation.} Through extensive simulation studies, we show that CEDM improves conditional expectation estimation relative to the compared baselines, and that CEDMI combines well-controlled size with favourable power relative to the competing methods. A flow cytometry application further illustrates how the method can be used to assess disputed edges within a scientifically motivated working graph.
\end{enumerate}

The remainder of the paper is organised as follows. Section~2 introduces CEDM and its training objective. Section~3 develops $\doop$-distribution sampling and the CEDMI edge test. Section~4 presents theoretical guarantees for observational and interventional sampling, as well as inference validity. Section~5 reports numerical studies on simulated graphs and a flow cytometry signalling network. Finally, Section~6 concludes the article and discusses future directions. Additional simulations and technical proofs are provided in the Supplementary Materials \citep{chen2026supp}.

\section{Causality-encoded diffusion models}

Let $\boldsymbol{X}=(\boldsymbol{X}_1^\top,\ldots,\boldsymbol{X}_p^\top)^\top$
be a collection of random vectors, where each node $j\in\mathcal{V}:=\{1,\ldots,p\}$
corresponds to a $d_j$-dimensional random vector
$\boldsymbol{X}_j\in\mathbb{R}^{d_j}$, and the total dimension is
$d=\sum_{j=1}^p d_j$. We assume that $\boldsymbol{X}$ is generated by structural causal models \citep{pearl2009causality} of the form 
\begin{equation}
	\label{eq:sem}
	\boldsymbol{X}_j
	= \boldsymbol{g}_j\bigl(\boldsymbol{X}_{\mathrm{pa}(j)}, \boldsymbol{U}_j\bigr),
	\quad j \in \mathcal{V},
\end{equation}
where $\mathrm{pa}(j)\subseteq \mathcal{V}\setminus\{j\}$ denotes the parent set
of node $j$ in a given directed acyclic graph (DAG), and
$\boldsymbol{X}_{\mathrm{pa}(j)} := (\boldsymbol{X}_k)_{k\in \mathrm{pa}(j)}$
collects the corresponding parent variables (concatenated into a single vector
when convenient). The functions $\boldsymbol{g}_j$ characterise the parent--child mechanisms,
and the noise vectors $\{\boldsymbol{U}_j\}_{j=1}^p$ are assumed to be mutually
independent.

\subsection{Causality-encoded diffusion process}

We next couple the structural causal model \eqref{eq:sem} with a diffusion-based
generative mechanism. Diffusion models synthesise data by learning to reverse a
stochastic process that progressively perturbs the data distribution with
Gaussian noise \citep{sohl2015deep,ho2020denoising,song2021scorebased}. We first recall the forward diffusion and its reverse-time
formulation, and then describe how to encode the causal structure \eqref{eq:sem}
into the reverse dynamics.

\paragraph{General diffusion process.}
Let $p_0$ denote the data distribution of $\boldsymbol{X}\in\mathbb{R}^{d}$ at
time $t=0$. The forward diffusion process gradually adds Gaussian noise via
\begin{equation}
	\label{eq:forward_process}
	\mathrm{d}\boldsymbol{X}(t)
	= -\frac{1}{2}\,\boldsymbol{X}(t)\,\mathrm{d}t
	+ \mathrm{d}\boldsymbol{W}(t),
	\qquad \boldsymbol{X}(0) \sim p_0,
\end{equation}
where $\boldsymbol{W}(t)$ is a standard Wiener process in $\mathbb{R}^{d}$. Let
$p_t$ denote the law of $\boldsymbol{X}(t)$. As $t\to\infty$, $p_t$ converges to
the standard Gaussian distribution; in practice, the process is run up to a
large but finite time horizon $T$.

Define the reverse-time process by $\boldsymbol{X}^{\leftarrow}(t)=\boldsymbol{X}(T-t)$.
The corresponding reverse SDE transports noise back to data and admits $p_0$ as
its terminal distribution:
$$\mathrm{d}\boldsymbol{X}^{\leftarrow}(t)
=
\left[
\frac{1}{2}\,\boldsymbol{X}^{\leftarrow}(t)
+ \nabla \log p_{T-t}\bigl(\boldsymbol{X}^{\leftarrow}(t)\bigr)
\right]\mathrm{d}t
+ \mathrm{d}\boldsymbol{\overline{W}}(t),
\qquad
\boldsymbol{X}^{\leftarrow}(0)\sim p_T,$$
where $\boldsymbol{\overline{W}}(t)$ is a time-reversed Wiener process and
$\nabla \log p_{T-t}$ is the score function at time $T-t$ \citep{anderson1982reverse}. Under mild regularity
conditions, the solution satisfies $\boldsymbol{X}^{\leftarrow}(T)\sim p_0$.

The diffusion model above learns the observational law $p_0$ (and the time-marginals $\{p_t\}_{t\in[0,T]}$) but is causality-agnostic: the forward SDE~\eqref{eq:forward_process} is graph-independent, and the reverse dynamics are parameterised by the joint score $\nabla \log p_t(\mathbf{x})$, so the update for $\boldsymbol{X}_j$ may depend on many coordinates rather than only on $\pa(j)$. It can therefore reproduce $p_0$ without distinguishing observationally equivalent graphs or identifying interventional laws such as $p_0(\boldsymbol{X}_{-j}\mid \doop(\boldsymbol{X}_j=\mathbf{x}_j))$. Our \emph{causality-encoded diffusion process} keeps the same forward SDE but constrains the reverse-time dynamics so that each block evolves conditional only on its parents in a specified DAG.

\paragraph{Causality-encoded diffusion process.}
Let $\mathcal{G}=(\mathcal{V},\mathcal{E})$ be a DAG with parent sets $\{\pa(j)\}_{j\in\mathcal{V}}$. If $p_0$ is Markov with respect to $\mathcal{G}$, then $p_0(\mathbf{x})=\prod_{j\in\mathcal{V}} p_0\bigl(\mathbf{x}_j \mid \mathbf{x}_{\pa(j)}\bigr),$ suggesting that each block be modelled conditionally on its parents. We enforce this structure throughout diffusion time by learning (or, in this idealised presentation, assuming access to) the conditional scores $\nabla_{\mathbf{x}'_j}\log p_t(\mathbf{x}'_j\mid \mathbf{x}_{\pa(j)})$, and constraining the reverse dynamics accordingly.

Let $\pi$ be any topological order of $\mathcal{G}$, so that $\pa(\pi(\ell)) \subseteq \{\pi(1),\ldots,\pi(\ell-1)\}$ for all $\ell$. In the resulting node-by-node reverse simulation, each node’s parents have already been generated and are treated as fixed inputs. We therefore propose the following DAG-respecting reverse-time sampler.

\begin{algorithm}[h]
	\caption{Causality-Encoded Diffusion on a DAG $\mathcal{G}$}
	\label{alg:dag_cedm_ideal}
	\SetAlgoLined
	\KwIn{
		DAG $\mathcal{G}=(\mathcal{V},\mathcal{E})$ with parent sets $\mathrm{pa}(j)$;\\
		exact conditional score functions
		$\mathbf{s}^{\star}_j(\mathbf{x}_j', \mathbf{x}_{\pa(j)}, t)
		= \nabla_{\mathbf{x}'_j}\log p_{t}\bigl(\mathbf{x}'_j \mid \mathbf{x}_{\pa(j)}\bigr)$
		for all $j\in\mathcal{V}$
		(with $\mathbf{s}^{\star}_j(\mathbf{x}_j', t)
		= \nabla_{\mathbf{x}'_j}\log p_{t}\bigl(\mathbf{x}'_j\bigr)$ for root nodes).
	}
	\KwOut{A sample $\mathbf{X}\in\mathbb{R}^{d}$}
	\For{$\ell=1,2,\dots,p$}{
		$j \gets \pi(\ell)$\;
		Evolve the $d_j$-dimensional reverse SDE
		\begin{equation}
			\label{eq:reversed_process_general}
			\mathrm{d}\boldsymbol{X}_j^{\leftarrow}(t)
			=
			\Bigl[
			\tfrac{1}{2}\,\boldsymbol{X}_j^{\leftarrow}(t)
			+
			\mathbf{s}^{\star}_j\bigl(
			\boldsymbol{X}_j^{\leftarrow}(t),
			\mathbf{X}_{\mathrm{pa}(j)},
			T-t
			\bigr)
			\Bigr]\mathrm{d}t
			+
			\mathrm{d}\boldsymbol{\overline{W}}_j(t),
		\end{equation}
		initialised at $\boldsymbol{X}_j^{\leftarrow}(0)\sim p_{T}(\cdot \mid \mathbf{X}_{\mathrm{pa}(j)})$,
		and set $\mathbf{X}_j := \boldsymbol{X}_j^{\leftarrow}(T)$\;
	}
	\Return{$\mathbf{X}$}\;
\end{algorithm}

Algorithm~\ref{alg:dag_cedm_ideal} differs from the standard diffusion sampler by replacing the \emph{joint} score $\nabla \log p_t(\mathbf{x})$ with \emph{local} conditional scores $\{\nabla_{\mathbf{x}'_j}\log p_t(\mathbf{x}'_j\mid \mathbf{x}_{\pa(j)})\}_{j=1}^p$ consistent with the DAG factorisation. Traversing nodes in topological order, we evolve for each $j$ a $d_j$-dimensional reverse SDE whose drift depends only on the current $\boldsymbol{X}_j^{\leftarrow}(t)$ and the already-sampled parent values $\mathbf{X}_{\pa(j)}$, with initialisation $\boldsymbol{X}_j^{\leftarrow}(0)\sim p_T(\cdot\mid \mathbf{X}_{\pa(j)})$ and output $\mathbf{X}_j=\boldsymbol{X}_j^{\leftarrow}(T)$. Under oracle conditional scores, this sampler recovers the observational factorisation exactly. This gives the sampler causal semantics and enables interventional sampling by freezing intervened nodes and propagating effects downstream through the DAG.

\subsection{Score estimations and training guidance}

In practice, the oracle conditional score functions
$\mathbf{s}^{\star}_j(\mathbf{x}'_j,\mathbf{x}_{\pa(j)},t)$ are not available and
must therefore be estimated from data. To this end, we consider the following
class of ReLU neural networks, denoted by $\mathcal{F}$:
\begin{equation*}
	\begin{aligned}
		&\mathcal{F}\!\left(M_t, W, \kappa, L, K\right) \\ := &\Bigl\{ \;
		\mathbf{s}(\mathbf{x}'_j, \mathbf{x}_{\pa(j)}, t)
		=\left(\boldsymbol{A}_L \sigma(\cdot)+\boldsymbol{b}_L\right) \circ \cdots \circ
		\left(\boldsymbol{A}_1\left[\mathbf{x}'_j, \mathbf{x}_{\pa(j)}^{\top}, t\right]^{\top}+\boldsymbol{b}_1\right)
		:\; \\
		& \boldsymbol{A}_i \in \mathbb{R}^{L_i \times L_{i+1}},\;
		\boldsymbol{b}_i \in \mathbb{R}^{L_{i+1}},\;
		\max_i L_i \leq W,\;
		\sup_{\mathbf{x}'_j,\,\mathbf{x}_{\pa(j)},\,t}\|\mathbf{s}(\mathbf{x}'_j, \mathbf{x}_{\pa(j)}, t)\|_{\infty} \leq M_t, \\
		& \max_{i}\bigl(\|\boldsymbol{A}_i\|_{\infty} \vee \|\boldsymbol{b}_i\|_{\infty}\bigr) \leq \kappa,\;
		\sum_{i=1}^L\bigl(\|\boldsymbol{A}_i\|_0+\|\boldsymbol{b}_i\|_0\bigr) \leq K
		\;\Bigr\},
	\end{aligned}
\end{equation*}
where $\sigma(\cdot)$ denotes the ReLU activation, $\|\cdot\|_{\infty}$ denotes the
maximum absolute entry, and $\|\cdot\|_{0}$ counts the number of non-zero entries.
The complexity of $\mathcal{F}$ is governed by the depth $L$, the maximal width $W$,
the magnitude and sparsity of the parameters $(\boldsymbol{A}_i,\boldsymbol{b}_i)$, and the
uniform bound $M_t$ on the network output.

We next examine the extent to which the class $\mathcal{F}$ can approximate the
(conditional) score functions arising in the reverse-time dynamics. 

\begin{definition}[H\"older ball]
	\label{def:holder_ball}
	Let $\beta = \lfloor \beta \rfloor + \gamma > 0$, with $\gamma \in [0,1)$. For a function $f: \mathbb{R}^r \rightarrow \mathbb{R}$, the H\"older ball of radius $B>0$ is defined as
	$$\mathcal{H}^{\beta}\left(\mathbb{R}^{r}, B\right) = \Bigl\{ f: \mathbb{R}^{r} \rightarrow \mathbb{R} \; \Bigl| \; \|f\|_{\mathcal{H}^{\beta}\left(\mathbb{R}^{r}\right)} < B \Bigr\},$$
	where the H\"older norm is given by
	$$\|f\|_{\mathcal{H}^{\beta}\left(\mathbb{R}^{r}\right)} := \max_{\|\mathbf{v}\|_{1} < \lfloor \beta \rfloor} \sup_{\mathbf{x}} \left|\partial^{\mathbf{v}} f(\mathbf{x})\right|
	+ \max_{\|\mathbf{v}\|_{1} = \lfloor \beta \rfloor} \sup_{\mathbf{x} \neq \mathbf{z}}
	\frac{\left|\partial^{\mathbf{v}} f(\mathbf{x}) - \partial^{\mathbf{v}} f(\mathbf{z})\right|}{\|\mathbf{x}-\mathbf{z}\|_{\infty}^{\gamma}}.$$
\end{definition}

\begin{assumption}
	\label{assumption:light_tail_assumption_data_stronger}
	For each $j\in\{1,\ldots,p\}$, assume there exist constants $C_{1,j}, C_{2,j} > 0$
	such that $f_j(\mathbf{x}_j,\mathbf{x}_{\pa(j)}) \geq C_{1,j}$ for all
	$(\mathbf{x}_j,\mathbf{x}_{\pa(j)})$, and the conditional density admits the form
	$$p(\mathbf{x}_j \mid \mathbf{x}_{\pa(j)})
	=
	\exp\!\bigl(-C_{2,j}\|\mathbf{x}_j\|^2\bigr)\, f_j(\mathbf{x}_j,\mathbf{x}_{\pa(j)}),$$
	where $f_j \in \mathcal{H}^{\beta}(\mathbb{R}^{d_j+r_j},B)$ for some $B>0$ and
	$\beta>1$, and $r_j := \sum_{\ell \in \pa(j)} d_{\ell}$.
\end{assumption}

Under Assumption~\ref{assumption:light_tail_assumption_data_stronger}, for each
$j\in\mathcal{V}$ we impose uniform lower (and implicit upper) bounds on $f_j$ to
ensure well-posed and statistically stable density estimation; see, for example,
\citet{wasserman2006all}. Closely related regularity conditions are also adopted
by \citet{fu2024unveil} in their analysis of (conditional) score-function
approximation.

\begin{proposition}
	\label{prop:score_approximation_stronger}
	Suppose Assumption~\ref{assumption:light_tail_assumption_data_stronger} holds.
	Then, for all sufficiently large $N$, there exist constants $C_{\sigma},C_{\alpha}>0$
	such that, with early-stopping time $t_0 = N^{-C_{\sigma}}$ and terminal time
	$T = C_{\alpha}\log N$, one can find $\mathbf{s}=\{\mathbf{s}_j\}_{j=1}^{p}$ with
	$\mathbf{s}_j \in \mathcal{F}(M_{t},W,\kappa,L,K)$ satisfying, for all
	$t\in[t_0,T]$,
	\begingroup
	\small
	\setlength{\jot}{1pt} 
	\begin{equation*}
		\begin{aligned}
			\mathbb{E}_{\mathbf{x} \sim p_{0}}\,
			\mathbb{E}_{ \mathbf{x}' \sim p_{t}(\cdot \mid \mathbf{x})}\!\left[
			\sum_{j=1}^p
			\left\|\mathbf{s}_j\!\left(\mathbf{x}'_j, \mathbf{x}_{\pa(j)}, t\right)
			-\nabla_{\mathbf{x}'_j} \log p_{t}\!\left(\mathbf{x}'_j \mid \mathbf{x}_{\pa(j)}\right)\right\|^2
			\right]
			=\mathcal{O}\!\left(\frac{1}{\sigma_{t}^2}\, N^{-\frac{2 \beta}{q}} (\log N)^{\beta +1} \right),
		\end{aligned}
	\end{equation*}
	\endgroup
	where $q = \max\{d_1+r_1,\ldots,d_p+r_p\}$ and $\sigma_t^2 = 1 - \exp(-t)$. Moreover, the network hyperparameters
	may be chosen so that $M_{t} = \mathcal{O}(\sqrt{\log N}/\sigma_{t}),
	W = \mathcal{O}(N \log^7 N),
	\kappa = \exp(\mathcal{O}(\log^4 N)),
	L = \mathcal{O}(\log^4 N),
	K = \mathcal{O}(N \log^9 N)$,
	where the $\mathcal{O}(\cdot)$ notation suppresses factors depending only on
	$C_\alpha$, $C_{\sigma}$, $\beta$, $B$, $\{d_j\}_{j=1}^p$, $\{C_{1,j}\}_{j=1}^{p}$,
	and $\{C_{2,j}\}_{j=1}^p$.
\end{proposition}

Proposition~\ref{prop:score_approximation_stronger} indicates that the network
class $\mathcal{F}$ is sufficiently expressive to approximate the conditional
score functions over a non-trivial time window, and thus has the potential to
serve as a practical estimator of $\{\mathbf{s}_j^\star\}_{j=1}^p$. Importantly,
the approximation error involves the factor $1/\sigma_t^2$, which deteriorates as
$t\downarrow 0$. This motivates the introduction of an early-stopping time
$t_0>0$. We now formalise the population risk and the empirical objective used
to fit the score networks.

Given a collection of score approximators $\mathbf{s}=\{\mathbf{s}_j\}_{j=1}^p$,
we seek to control the integrated score-matching risk
\begin{equation}
	\label{eq:loss_score_match}
	\begin{aligned}
		\mathcal{R}(\mathbf{s})
		:= \frac{1}{T-t_0}\int_{t_0}^{T}
		\mathbb{E}_{\mathbf{x}\sim p_0}\Biggl[
		\mathbb{E}_{\mathbf{x}'\sim p_{t}(\cdot \mid \mathbf{x})}\Biggl[
		\sum_{j=1}^p
		\left\|
		\mathbf{s}_j\!\left(\mathbf{x}'_j, \mathbf{x}_{\pa(j)}, t\right)
		-
		\nabla_{\mathbf{x}'_j}\log p_{t}\!\left(\mathbf{x}'_j \mid \mathbf{x}_{\pa(j)}\right)
		\right\|^2
		\Biggr]
		\Biggr]\,\mathrm{d}t .
	\end{aligned}
\end{equation}

In practice, the conditional score
$\nabla_{\mathbf{x}'_j}\log p_{t}\bigl(\mathbf{x}'_j \mid \mathbf{x}_{\pa(j)}\bigr)$
is unknown. We therefore minimise an equivalent objective based on denoising
score matching \citep{vincent2011connection}:
\begin{equation}
	\label{eq:equivalent_loss}
	\begin{aligned}
		\mathcal{L}(\mathbf{s})
		:= \frac{1}{T-t_0}\int_{t_0}^{T}
		\mathbb{E}_{\mathbf{x}\sim p_0}\Biggl[
		\mathbb{E}_{\mathbf{x}'\sim \mathcal{N}\!\left(\alpha_t \mathbf{x}, \sigma_t^2 \boldsymbol{I}\right)}\Biggl[
		\sum_{j=1}^p
		\left\|
		\mathbf{s}_j\!\left(\mathbf{x}'_j, \mathbf{x}_{\pa(j)}, t\right)
		-
		\nabla_{\mathbf{x}'_j}\log \phi_{t}\!\left(\mathbf{x}'_j \mid \mathbf{x}_j\right)
		\right\|^2
		\Biggr]
		\Biggr]\,\mathrm{d}t .
	\end{aligned}
\end{equation}
Here $\phi_t$ denotes the Gaussian transition kernel induced by the forward
process \eqref{eq:forward_process}, and
$$\nabla_{\mathbf{x}'_j}\log \phi_{t}\!\left(\mathbf{x}'_j \mid \mathbf{x}_j\right)
= -\frac{\mathbf{x}'_j - \alpha_t \mathbf{x}_{j}}{\sigma_{t}^2},
\qquad
\alpha_t = \exp(-t/2),\quad \sigma_t^2 = 1-\alpha_t^2 .$$

\begin{lemma}
	\label{lemma:loss_equivalence}
	The difference $\mathcal{R}(\mathbf{s})-\mathcal{L}(\mathbf{s})$ does not depend on $\mathbf{s}$.
\end{lemma}

Lemma~\ref{lemma:loss_equivalence} shows that $\mathcal{L}(\mathbf{s})$ agrees
with $\mathcal{R}(\mathbf{s})$ up to an additive constant that is independent of
$\mathbf{s}$. Consequently, minimising the tractable objective
\eqref{eq:equivalent_loss} is equivalent to minimising the intractable risk
\eqref{eq:loss_score_match}, thereby justifying the use of denoising score
matching for training.

In implementation, \eqref{eq:equivalent_loss} is approximated using i.i.d.
samples $\{\mathbf{X}_i\}_{i=1}^n \sim p_0$, which replace the population
expectation over $\boldsymbol{X}$ by its empirical analogue. The
causality-encoded guidance is thus obtained by minimising the empirical loss
\begin{equation}
	\label{eq:empirical_loss}
	\widehat{\mathcal{L}}(\mathbf{s})
	=\frac{1}{n}\sum_{i=1}^n \ell\!\left(\mathbf{X}_i, \mathbf{s}\right),
\end{equation}
where
\begin{equation}
	\ell\!\left(\mathbf{x}, \mathbf{s}\right)
	=\frac{1}{T-t_0}\int_{t_0}^{T}
	\mathbb{E}_{\mathbf{x}'\sim \mathcal{N}\!\left(\alpha_t \mathbf{x}, \sigma_t^2 \boldsymbol{I}\right)}\Biggl[
	\sum_{j=1}^p
	\left\|
	\mathbf{s}_j\!\left(\mathbf{x}'_j, \mathbf{x}_{\pa(j)}, t\right)
	-\nabla_{\mathbf{x}'_j}\log \phi_{t}\!\left(\mathbf{x}'_j \mid \mathbf{x}_j\right)
	\right\|^2
	\Biggr]\,\mathrm{d}t .
\end{equation}

Let
$$\widehat{\mathbf{s}}
\in \arg\min_{\mathbf{s}\in \mathcal{F}^p}
\frac{1}{n}\sum_{i=1}^n \ell\!\left(\mathbf{X}_i, \mathbf{s}\right)$$
denote a minimiser of \eqref{eq:empirical_loss}. We then implement the
reverse-time dynamics \eqref{eq:reversed_process_general} by replacing the
population scores with $\widehat{\mathbf{s}}$:
\begin{equation}
	\label{eq:reversed_process_empirical}
	\mathrm{d}\widehat{\boldsymbol{X}}_j^{\leftarrow}(t)
	=
	\Bigl[
	\tfrac{1}{2}\,\widehat{\boldsymbol{X}}_j^{\leftarrow}(t)
	+
	\widehat{\mathbf{s}}_j\bigl(
	\widehat{\boldsymbol{X}}_j^{\leftarrow}(t),
	\mathbf{X}_{\pa(j)},
	T-t
	\bigr)
	\Bigr]\mathrm{d}t
	+
	\mathrm{d}\boldsymbol{\overline{W}}_j(t),
\end{equation}
initialised at $\widehat{\boldsymbol{X}}_j^{\leftarrow}(0)\sim \mathcal{N}(\mathbf{0},\boldsymbol{I})$,
and we set $\mathbf{X}_j := \widehat{\boldsymbol{X}}_j^{\leftarrow}(T-t_0)$.

\begin{proposition}
	\label{propo:score_risk_bound_stronger}
	Suppose the conditions of Proposition~\ref{prop:score_approximation_stronger} hold,
	and the score networks are chosen as in Proposition~\ref{prop:score_approximation_stronger}.
	With the network-size parameter $N = n^{\frac{q}{q+2\beta}}$, early-stopping time $t_0<1$,
	and terminal time $T=\mathcal{O}(\log n)$, we have
	$$\mathbb{E}_{\{\mathbf{X}_i\}_{i=1}^n}\!\left[\mathcal{R}(\widehat{\mathbf{s}})\right]
	=
	\mathcal{O}\!\left(
	\log\!\frac{1}{t_0}\cdot n^{-\frac{2\beta}{q+2\beta}} \log^{\max(17,\beta+1)} n
	\right).$$
\end{proposition}

Proposition~\ref{propo:score_risk_bound_stronger} formalises the central statistical advantage of encoding the DAG. The rate depends on $q=\max_{j\in\mathcal{V}}(d_j+r_j)$, the largest dimension of any node together with its parents, rather than on the ambient dimension $d=\sum_{j=1}^p d_j$. The estimation burden is therefore local: each score network learns a parent--child conditional relation instead of a full joint score. When the graph is sparse, $q$ can be substantially smaller than $d$, yielding a corresponding reduction in effective nonparametric complexity. The same locality carries through to the distribution-recovery bound in Theorem~\ref{theorem:total_variation_bound}.

\section{Interventional sampling and edge inference}

The fitted local-score system serves two purposes. First, it yields Monte Carlo samples from interventional laws under a specified DAG. Second, the same generator can be used to simulate null distributions for targeted edge tests within a working graph. We develop these uses in this section.

\subsection{CEDM for \texorpdfstring{$\doop$}{do}-interventional sampling}
\label{subsec:cedm-do}

We briefly recall do-interventions in structural causal models~\citep{pearl2009causality}. For $\mathcal{S}\subseteq\mathcal{V}$ and $\mathbf{x}_{\mathcal{S}}^\star$, $\doop(\boldsymbol{X}_{\mathcal{S}}=\mathbf{x}_{\mathcal{S}}^\star)$ sets $\boldsymbol{X}_j=\mathbf{x}_j^\star$ for all $j\in\mathcal{S}$ whilst leaving the remaining structural equations unchanged. The resulting \emph{do}-distribution
$p(\mathbf{x}\mid \doop(\boldsymbol{X}_{\mathcal{S}}=\mathbf{x}_{\mathcal{S}}^\star))$
generally differs from the observational conditional
$p(\mathbf{x}\mid \boldsymbol{X}_{\mathcal{S}}=\mathbf{x}_{\mathcal{S}}^\star)$.
Under a DAG, it admits the truncated factorisation~\citep{pearl2009causality}
\begin{equation}
	p(\mathbf{x} \mid \doop(\boldsymbol{X}_\mathcal{S} = \mathbf{x}_\mathcal{S}^\star))
	=
	\delta(\mathbf{x}_\mathcal{S} - \mathbf{x}_\mathcal{S}^\star)
	\prod_{j \notin \mathcal{S}} p_j\Bigl(
	\mathbf{x}_j \,\Big|\,
	\boldsymbol{X}_{\pa(j)\cap\mathcal{S}}=\mathbf{x}_{\pa(j)\cap\mathcal{S}}^\star,\;
	\boldsymbol{X}_{\pa(j)\setminus\mathcal{S}}=\mathbf{x}_{\pa(j)\setminus\mathcal{S}}
	\Bigr),
	\label{eq:truncated_factorization}
\end{equation}
where $\delta(\mathbf{x}_\mathcal{S}-\mathbf{x}_\mathcal{S}^\star)$ is a point mass at $\mathbf{x}_\mathcal{S}^\star$. This corresponds to cutting all incoming edges into $\boldsymbol{X}_{\mathcal{S}}$ whilst preserving the remaining local mechanisms: ancestors retain their marginal laws, whereas descendants may change through modified upstream inputs.

Algorithm~\ref{alg:dag_cedm_do} implements the same semantics in reverse diffusion by freezing $\boldsymbol{X}_{\mathcal{S}}$ at $\mathbf{x}_{\mathcal{S}}^\star$ throughout the reverse-time simulation and evolving only the non-intervened nodes using the causality-encoded conditional scores, yielding samples from the target do-distribution when these scores are well-approximated.

\begin{algorithm}[ht]
	\caption{CEDM sampling under intervention $\doop(\boldsymbol{X}_{\mathcal{S}} = \mathbf{x}_{\mathcal{S}}^\star)$}
	\label{alg:dag_cedm_do}
	\SetAlgoLined
	\KwIn{
		DAG $\mathcal{G} = (\mathcal{V},\mathcal{E})$ with parent sets $\{\mathrm{pa}(j)\}_{j=1}^p$ and a causal order $\pi$; CEDM networks $\widehat{\mathbf{s}} = \{\widehat{\mathbf{s}}_j\}_{j=1}^p$; intervention set $\mathcal{S} \subseteq \mathcal{V}$ and target value $\mathbf{x}_{\mathcal{S}}^\star$.
	}
	\KwOut{A sample $\mathbf{X}$ with law $p(\cdot \mid \doop(\boldsymbol{X}_{\mathcal{S}} = \mathbf{x}_{\mathcal{S}}^\star))$}
	
	\For{$\ell = 1,\ldots,p$}{
		$j \gets \pi(\ell)$\;
		\eIf{$j \notin \mathcal{S}$}{
			Evolve the reverse SDE in \eqref{eq:reversed_process_empirical} and set
			$\mathbf{X}_j \gets \widehat{\boldsymbol{X}}^{\leftarrow}_j(T-t_0)$\;
		}{
			$\mathbf{X}_j \gets \mathbf{x}_j^\star$\;
		}
	}
	\Return{$\mathbf{X}$}\;
\end{algorithm}

Algorithm~\ref{alg:dag_cedm_do} yields Monte Carlo samples from interventional laws on a DAG. With samples from $p(\cdot \mid \doop(\boldsymbol{X}_{\mathcal{S}}=\mathbf{x}_{\mathcal{S}}^\star))$, any causal functional expressible as a $\doop$-expectation can be estimated by sample averaging, and average or heterogeneous effects can be obtained by contrasting interventions. Such samples also facilitate distributional and decision-relevant summaries, including quantiles and tail risks, without requiring additional closed-form identification beyond the DAG. We employ $\doop$-sampling as a generative building block for inference: in the sequel, null distributions are simulated by intervening on a candidate graph and assessing conditional dependence.

\subsection{CEDM-based inference}

To preclude redundant parents, we assume \emph{causal minimality} for the SCM~\eqref{eq:sem}: for every $j\in\mathcal{V}$ and each $k\in\pa(j)$, the mechanism $\boldsymbol{g}_j$ depends non-trivially on its $k$-th parental argument, i.e.\ there exist values of $\mathbf{X}_{\pa(j)\setminus\{k\}}$ and $\mathbf{U}_j$ such that the map
$\mathbf{x}_k \mapsto \boldsymbol{g}_j(\mathbf{x}_k,\mathbf{X}_{\pa(j)\setminus\{k\}},\mathbf{U}_j)$
is not constant.

We now test for directed edges. Let $\mathcal{H}\subseteq\{(k,j):k\neq j\}$ be a set of hypothesised arrows $k\to j$, let $\mathcal{E}$ denote the unknown true edge set, and let $\mathcal{E}'$ be a working edge set satisfying $\mathcal{E}\subseteq \mathcal{E}'\cup\mathcal{H}$ (e.g.\ $\mathcal{E}'$ obtained from structure learning, with $\mathcal{H}$ collecting additional edges to assess). Write $\pa'(j)$ for the parent set of node $j$ in the working graph $(\mathcal{V},\mathcal{E}')$. Our goal is to determine whether any edge in $\mathcal{H}$ is present, leading to
\begin{equation}\label{equation:test}
	H_0:\ \mathcal{E} \subseteq \mathcal{E}' \setminus \mathcal{H}
	\quad \text{versus} \quad
	H_a:\ \mathcal{E} \not\subseteq \mathcal{E}' \setminus \mathcal{H}.
\end{equation}
Under $H_0$, all edges in $\mathcal{H}$ are absent; rejecting $H_0$ implies that at least one hypothesised edge belongs to $\mathcal{E}$.

We first consider a single candidate edge $(k,j)\in\mathcal{H}$. To encode the null $H_0:(k,j)\notin\mathcal{E}$, we delete $(k,j)$ from the working graph $\mathcal{E}'$ and use the CEDM do-sampler to resample $\boldsymbol{X}_j$ under this modified graph while holding $(\boldsymbol{X}_k,\boldsymbol{X}_{\pa'(j)\setminus\{k\}})$ at their observed values. Any remaining conditional dependence between $\boldsymbol{X}_j$ and $\boldsymbol{X}_k$ given $\boldsymbol{X}_{\pa'(j)\setminus\{k\}}$ is quantified by a dependence measure $\xi(\cdot)$ and calibrated via its CEDM resampling distribution. Algorithm~\ref{alg:cedmi_single} summarises the resulting single-edge CEDMI test.

\begin{algorithm}[ht]
	\caption{CEDMI for a single edge}
	\label{alg:cedmi_single}
	\SetAlgoLined
	\KwIn{
		Dataset $\mathcal{D}=\{\mathbf{X}_{i}\}_{i=1}^n$; working edge set $\mathcal{E}'$ with parent sets $\pa'(j)$; hypothesis $\mathcal{H}=\{(k,j)\}$; CEDM networks $\widehat{\mathbf{s}}=\{\widehat{\mathbf{s}}_j\}_{j=1}^p$ trained under the null graph with edge set $\mathcal{E}'\setminus\mathcal{H}$; Monte Carlo size $M_{\mathrm{mc}}$; conditional dependence measure $\xi_n(\mathbf{X}_{\cdot,j},\mathbf{X}_{\cdot,k},\mathbf{X}_{\cdot,\pa'(j)\setminus\{k\}})$ for testing $\boldsymbol{X}_j \perp \boldsymbol{X}_k \mid \boldsymbol{X}_{\pa'(j)\setminus\{k\}}$.
	}
	\KwOut{A $p$-value for the null hypothesis $H_0:(k,j)\notin\mathcal{E}$.}
	
	Compute the observed statistic
	$\widehat{\xi} := \xi_n(\mathbf{X}_{\cdot,j},\mathbf{X}_{\cdot,k},\mathbf{X}_{\cdot,\pa'(j)\setminus\{k\}})$\;
	
	Define the intervention set $\mathcal{S} := \{k\}\cup\bigl(\pa'(j)\setminus\{k\}\bigr)$\;
	
	\For{$m = 1,2,\dots,M_{\mathrm{mc}}$}{
		\For{$i = 1,2,\dots,n$}{
			Set the target value
			$\mathbf{x}_{\mathcal{S}}^\star := \bigl(\mathbf{X}_{i,k},\,\mathbf{X}_{i,\pa'(j)\setminus\{k\}}\bigr)$\;
			
			Run Algorithm~\ref{alg:dag_cedm_do} on the graph with edge set $\mathcal{E}'\setminus\mathcal{H}$ under the intervention
			$\doop(\boldsymbol{X}_{\mathcal{S}}=\mathbf{x}_{\mathcal{S}}^\star)$,
			and record the synthetic draw $\mathbf{X}_{i,j}^{(m)}$ for node $j$\;
		}
		Form $\mathbf{X}_j^{(m)} := (\mathbf{X}_{1,j}^{(m)},\ldots,\mathbf{X}_{n,j}^{(m)})^\top$\;
		
		Compute the $m$-th null statistic
		$\widehat{\xi}^{(m)} := \xi_n(\mathbf{X}_j^{(m)},\mathbf{X}_{\cdot,k},\mathbf{X}_{\cdot,\pa'(j)\setminus\{k\}})$\;
	}
	
	Compute the Monte Carlo $p$-value
	$p := \dfrac{1+\sum_{m=1}^{M_{\mathrm{mc}}}\mathbf{1}\{\widehat{\xi}^{(m)}\geq \widehat{\xi}\}}{1+M_{\mathrm{mc}}}$\; 
\end{algorithm}

In practice, to safeguard empirical size, one may add a simple diagnostic by comparing the empirical law of $(\mathbf{X}_{\cdot,j},\mathbf{X}_{\cdot,\pa'(j)\setminus\{k\}})$ with samples from the fitted diffusion model, e.g.\ via an MMD two-sample test~\citep{gretton2012kernel}. Algorithm~\ref{alg:cedmi_single} extends immediately to $\mathcal{H}$: compute a $p$-value for each $(k,j)\in\mathcal{H}$ and apply a standard multiple-testing correction (FWER or FDR)~\citep{holm1979simple,benjamini1995controlling}. For a single candidate edge, Algorithm~\ref{alg:cedmi_single} therefore entails one CEDM fit under the edge-deleted null graph $\mathcal{E}'\setminus\{(k,j)\}$ and for several hypotheses, the fit is repeated once per null graph under consideration.  In some applications, it is preferable to aggregate several edge hypotheses into a single joint test, since any violated edge in $\mathcal{H}$ manifests as a failure of an appropriate conditional independence relation.

To instantiate Algorithm~\ref{alg:cedmi_single}, we require a \emph{scalar} statistic that
vanishes under the conditional independence relation
$\boldsymbol{X}_j \perp \boldsymbol{X}_k \mid \boldsymbol{X}_{\mathrm{pa}'(j)\setminus\{k\}}$.
We obtain such a statistic by combining the conditional-dependence construction of
\citet{azadkia2021simple} with the multi-output extension of the Azadkia--Chatterjee
coefficient introduced by \citet{ansari2022simple}.

Let $\boldsymbol{X}_j \in \mathbb{R}^{d_j}$, $\boldsymbol{X}_k \in \mathbb{R}^{d_k}$, and
$\boldsymbol{X}_{\pa'(j)\setminus \{k\}}\in\mathbb{R}^{r_j}$.
When $d_j=1$, \citet{azadkia2021simple} introduced the following coefficient for measuring
the conditional dependence of $X_j$ on $\boldsymbol{X}_k$ given $\boldsymbol{X}_{\pa'(j)\setminus \{k\}}$:
\begin{equation}
	\xi(X_j, \boldsymbol{X}_k, \boldsymbol{X}_{\pa'(j)\setminus \{k\}})
	:=\frac{\int_{\mathbb{R}} \mathbb{E}(\operatorname{var}(P(X_j \geq \mathrm{x}_j \mid \boldsymbol{X}_k, \boldsymbol{X}_{\pa'(j)\setminus \{k\}}) \mid \boldsymbol{X}_{\pa'(j)\setminus \{k\}})) \mathrm{d} p_j(\mathrm{x}_j)}{\int_{\mathbb{R}} \mathbb{E}\left(\operatorname{var}\left(\mathbf{1}_{\{X_j \geq \mathrm{x}_j\}} \mid \boldsymbol{X}_{\pa'(j)\setminus \{k\}}\right)\right) \mathrm{d} p_j(\mathrm{x}_j)}.
\end{equation}
This measure has attracted substantial attention in recent years
\citep{huang2022kernel,shi2022power,lin2023boosting,dette2025simple}.
When $\pa'(j)\setminus \{k\}=\emptyset$ and $\boldsymbol{X}_j$ is multivariate,
\citet{ansari2022simple} proposed the following multi-output extension:
\begin{equation}
	\xi(\boldsymbol{X}_j,\boldsymbol{X}_k)
	\;:=\;
	1-\frac{
		d_j -\sum_{\ell=1}^{d_j} \xi\!\bigl(\boldsymbol{X}_{j,\ell},( \boldsymbol{X}_k, \boldsymbol{X}_{j,\ell-1},\ldots,\boldsymbol{X}_{j,1} )\bigr)
	}{
		d_j-\sum_{\ell=2}^{d_j} \xi\!\bigl(\boldsymbol{X}_{j,\ell},(\boldsymbol{X}_{j,\ell-1},\ldots,\boldsymbol{X}_{j,1} )\bigr)
	},
	\qquad \text{with }\ \xi(\boldsymbol{X}_{j,1},\emptyset):=0 .
	\label{eq:T_multi_output}
\end{equation}
Intuitively, $\xi(\boldsymbol{X}_j,\boldsymbol{X}_k)$ aggregates the \emph{incremental} predictability of
each coordinate $\boldsymbol{X}_{j,\ell}$ from $\boldsymbol{X}_k$, beyond the information already contained in
$(\boldsymbol{X}_{j,\ell-1},\ldots,\boldsymbol{X}_{j,1})$, whilst remaining fully rank-based and tuning-free.

Motivated by the information-gain perspective of Example~3.9 in \citet{ansari2022simple}, we introduce the following multivariate conditional dependence coefficient (MCODEC). We assume that $\boldsymbol{X}_j$ is not perfectly dependent on $\boldsymbol{X}_{\pa'(j)\setminus \{k\}}$, so that the normalisation below is well defined:
\begin{equation}
	\xi(\boldsymbol{X}_j,\boldsymbol{X}_k, \boldsymbol{X}_{\pa'(j)\setminus \{k\}})
	\;:=\;
	\frac{
		\xi(\boldsymbol{X}_j,(\boldsymbol{X}_{\pa'(j)\setminus \{k\}},\boldsymbol{X}_k))-\xi(\boldsymbol{X}_j,\boldsymbol{X}_{\pa'(j)\setminus \{k\}})
	}{
		1-\xi(\boldsymbol{X}_j,\boldsymbol{X}_{\pa'(j)\setminus \{k\}})
	}.
	\label{eq:mcodec_conditional_population}
\end{equation}
The numerator may be viewed as the additional predictability of $\boldsymbol{X}_j$ gained by
including $\boldsymbol{X}_k$ beyond $\boldsymbol{X}_{\pa'(j)\setminus \{k\}}$, and the denominator rescales this gain by the 
residual unpredictability of $\boldsymbol{X}_j$ given $\boldsymbol{X}_{\pa'(j)\setminus \{k\}}$, thereby producing a quantity
in $[0,1]$ with a clear operational interpretation.

With the axioms and information-gain properties established in \citet{ansari2022simple}, the
proposed MCODEC inherits the following desirable guarantees.

\begin{lemma}
	\label{lemma:mcodec_property}
	The multivariate conditional dependence coefficient defined in \eqref{eq:mcodec_conditional_population} when $\boldsymbol{X}_j$ is not perfectly dependent on $\boldsymbol{X}_{\pa'(j)\setminus \{k\}}$ satisfies the following properties:
    \begin{enumerate}[(a)]
        \item $0\le \xi(\boldsymbol{X}_j,\boldsymbol{X}_k, \boldsymbol{X}_{\pa'(j)\setminus \{k\}})\le 1$,
        \item $\xi(\boldsymbol{X}_j,\boldsymbol{X}_k, \boldsymbol{X}_{\pa'(j)\setminus \{k\}})=0$ if and only if $\boldsymbol{X}_j\perp \boldsymbol{X}_k\mid \boldsymbol{X}_{\pa'(j)\setminus \{k\}}$,
        \item $\xi(\boldsymbol{X}_j,\boldsymbol{X}_k, \boldsymbol{X}_{\pa'(j)\setminus \{k\}})=1$ if and only if $\boldsymbol{X}_j$ is perfectly dependent on $(\boldsymbol{X}_{\pa'(j)\setminus \{k\}},\boldsymbol{X}_k)$.
    \end{enumerate}
\end{lemma}

Lemma~\ref{lemma:mcodec_property} establishes that MCODEC is well matched to our edge test: it is in $[0,1]$ and vanishes \emph{if and only if} $\boldsymbol{X}_j \perp \boldsymbol{X}_k \mid \boldsymbol{X}_{\pa'(j)\setminus\{k\}}$. This bounded, rank-based, tuning-free construction yields stable Monte Carlo calibration in Algorithm~\ref{alg:cedmi_single}.

Given i.i.d.\ samples $\{(\mathbf{X}_{i,j},\mathbf{X}_{i,k}, \mathbf{X}_{i,\pa'(j) \setminus \{k\}})\}_{i=1}^n$, let $\xi_n$ denote the
estimator of \eqref{eq:T_multi_output} in \citet{ansari2022simple}. We define the empirical MCODEC estimator by
\begin{equation}
	\xi_n(\mathbf{X}_{\cdot,j},\mathbf{X}_{\cdot,k}, \mathbf{X}_{\cdot,\pa'(j) \setminus \{k\}})
	\;:=\;
	\frac{
		\xi_n(\mathbf{X}_{\cdot,j},(\mathbf{X}_{\cdot,\pa'(j) \setminus \{k\}},\mathbf{X}_{\cdot,k}))-\xi_n(\mathbf{X}_{\cdot,j},\mathbf{X}_{\cdot,\pa'(j) \setminus \{k\}})
	}{
		1-\xi_n(\mathbf{X}_{\cdot,j},\mathbf{X}_{\cdot,\pa'(j) \setminus \{k\}})
	}.
	\label{eq:mcodec_conditional_sample}
\end{equation}

Although the population coefficient in~\eqref{eq:mcodec_conditional_population} lies in $[0,1]$, the empirical ratio in~\eqref{eq:mcodec_conditional_sample} is not finite-sample range constrained; accordingly, Monte Carlo null statistics can occasionally be slightly negative even though the target parameter is nonnegative. As a direct consequence of Theorem~3.1 in \citet{ansari2022simple}, the proposed estimator is strongly consistent.

\begin{lemma}
	\label{lemma:mcodec_consistency}
	Assume $\boldsymbol{X}_j$ is not perfectly dependent on $\boldsymbol{X}_{\pa'(j)\setminus \{k\}}$; then $\lim_{n \to \infty} \xi_n \rightarrow \xi$ almost surely.
\end{lemma}

Lemma~\ref{lemma:mcodec_consistency} underpins power: under any fixed alternative with $\xi>0$, strong consistency yields $\xi_n \to \xi>0$ a.s., so $\xi_n$ separates from the null target $0$ and the Monte Carlo p-value in Algorithm~\ref{alg:cedmi_single} converges to $0$. In our experiments, we use $\xi_n(\mathbf{X}_{\cdot,j},\mathbf{X}_{\cdot,k}, \mathbf{X}_{\cdot,\mathrm{pa}'(j)\setminus\{k\}})$ and
$\xi_n(\mathbf{X}_j^{(m)},\mathbf{X}_{\cdot,k},\mathbf{X}_{\cdot,\mathrm{pa}'(j)\setminus\{k\}})$
as the observed and synthetic statistics in Algorithm~\ref{alg:cedmi_single}. 

  \section{Theoretical guarantees}

We now establish theoretical guarantees for causality-encoded diffusion models, beginning with a finite-sample bound for the learned observational law and then extending the result to $\doop$-interventional distributions. Finally, we provide type~I error control for the edge test. Throughout this section, we use $p_0$ to denote the observational distribution at time $0$, and we write $\widehat{p}_{t_0}$ for the distribution induced by the learned CEDM sampler with early stopping at diffusion time $t_0$.

\begin{theorem}
	\label{theorem:total_variation_bound}
	Suppose Assumption~\ref{assumption:light_tail_assumption_data_stronger} holds. Assume moreover that there exists a constant $C>0$ such that, for every $j\in\{1,\ldots,p\}$,
	$$\mathrm{KL}\!\left(p_0(\cdot \mid \mathbf{x}_{\pa(j)})\,\big\|\,\mathcal{N}(\mathbf{0},\boldsymbol{I}_{d_j})\right)\le C
	\quad\text{for all }\mathbf{x}_{\pa(j)}\in\mathbb{R}^{r_j}\ \text{when }\pa(j)\neq\emptyset,$$
	and, for root nodes, $\mathrm{KL}\!\left(p_0(\cdot)\,\big\|\,\mathcal{N}(\mathbf{0},\boldsymbol{I}_{d_j})\right)\le C$. Taking the early-stopping time $t_0 = n^{-\frac{4 \beta}{q+2\beta}-1}$ and the terminal time $T = \frac{4 \beta}{q+2\beta} \log n$, it holds that
	\begin{equation*}
		\E_{\{\mathbf{X}_i\}_{i=1}^n}\!\left[\operatorname{TV}\!\left(p_0,\widehat{p}_{t_0}\right)\right]
		= \mathcal{O}\!\left(n^{-\frac{\beta}{q+2\beta}}\log^{\max(\frac{19}{2},\frac{\beta+3}{2})} n\right).
	\end{equation*}
\end{theorem}

Theorem~\ref{theorem:total_variation_bound} provides the clearest statement of the paper's statistical claim. The recovery rate is driven by $q$, the maximum dimension of a node together with its parents. For sparse graphs this can be far smaller than the ambient dimension $d$, so the DAG acts as an explicit form of structural dimension reduction.

We next extend this guarantee from the observational law to interventional distributions. In particular, the accuracy of a causality-encoded diffusion model for sampling from $p(\cdot \mid \doop(\boldsymbol{X}_{\mathcal{S}}=\mathbf{x}_{\mathcal{S}}^\star))$
depends crucially on whether the observational data place sufficient mass in the region induced by fixing
$\boldsymbol{X}_{\mathcal{S}}=\mathbf{x}_{\mathcal{S}}^\star$. In the literature, this phenomenon is commonly described as \emph{distribution shift}: it governs the extent to which information learned from the training distribution transfers to the target intervention value; see \citet{yuan2023reward,fu2024unveil} in the diffusion model context.

To make this notion quantitative, for $j \not \in \mathcal{S}$ we define 
\begingroup
\small
\setlength{\jot}{1pt} 
\begin{equation}
	\label{eq:coeff_shift_node_wise}
	\begin{aligned}
		& \mathcal{T}_j(\mathbf{s}, \mathbf{x}_{\mathcal{S}}^\star) \\ := &\sqrt{ \frac{\int_{t_0}^T \E_{\mathbf{x}_{\pa(j) \setminus \mathcal{S}} \sim p(\cdot \mid \doop(\boldsymbol{X}_{\mathcal{S}}=\mathbf{x}_{\mathcal{S}}^\star))} \left[\E_{\mathbf{x}'_j \sim p_t(\cdot \mid \mathbf{x}_{\pa(j) \setminus \mathcal{S}},\mathbf{x}_{\pa(j) \cap \mathcal{S}}^\star)} \left[\Delta^2_j(\mathbf{x}'_j, \mathbf{x}_{\pa(j) \setminus \mathcal{S}},\mathbf{x}_{\pa(j) \cap \mathcal{S}}^\star ,t)\right]\right]\mathrm{~d} t}{\sum_{\ell = 1}^p \int_{t_0}^T \E_{\mathbf{x}_{\pa(\ell)}} \left[\E_{\mathbf{x}'_\ell \sim p_t(\cdot \mid \mathbf{x}_{\pa(\ell)})} \left[\Delta^2_\ell(\mathbf{x}'_\ell, \mathbf{x}_{\pa(\ell)},t)\right]\right]\mathrm{~d} t} },
	\end{aligned}
\end{equation}  
\endgroup
where 
\begin{equation}
	\label{eq:score_mismatch}
	\begin{aligned}
		& \Delta_j(\mathbf{x}'_j, \mathbf{x}_{\pa(j) \setminus \mathcal{S}},\mathbf{x}_{\pa(j) \cap \mathcal{S}}^\star ,t) \\:= &\left\|\mathbf{s}_j\left(\mathbf{x}'_j, \mathbf{x}_{\pa(j) \setminus \mathcal{S}},\mathbf{x}_{\pa(j) \cap \mathcal{S}}^\star ,t\right)-\nabla_{\mathbf{x}'_j} \log p_{t}\left(\mathbf{x}'_j \mid \mathbf{x}_{\pa(j) \setminus \mathcal{S}},\mathbf{x}_{\pa(j) \cap \mathcal{S}}^\star\right)\right\|,
	\end{aligned}
\end{equation}
and $\mathcal{T}_j(\mathbf{s}, \mathbf{x}_{\mathcal{S}}^\star):=0$ for $j \in \mathcal{S}$. This leads to the following class-restricted distribution shift coefficient:
\begin{equation}
	\begin{aligned}
		\mathcal{T}(\mathbf{x}_{\mathcal{S}}^\star) = \max \left( \sup_{\mathbf{s} \in \mathcal{F}^p} \max_{j \in \{1,\dots,p\}} \mathcal{T}_j(\mathbf{s}, \mathbf{x}_{\mathcal{S}}^\star),1\right).
	\end{aligned}
\end{equation}
Intuitively, $\mathcal{T}(\mathbf{x}_{\mathcal{S}}^\star)$ records how much the intervention can amplify score mismatch relative to the aggregate training error. When the intervention targets remain in regions with adequate observational support, one expects this factor to stay controlled. If $\mathbf{x}_{\mathcal{S}}^\star$ moves the system into a weakly supported region of covariate space, then even a well-trained model may incur a larger multiplicative constant under that intervention.

\begin{theorem}
	\label{theorem:total_variation_bound_do_intervention}
	Suppose Assumptions in Theorem \ref{theorem:total_variation_bound} hold. Then for any fixed $\mathbf{x}_{\mathcal{S}}^\star$, it holds that
	\begingroup
	\small
	\setlength{\jot}{1pt} 
	\begin{equation*}
		\E_{\{\mathbf{X}_i\}_{i=1}^n} \left[\operatorname{TV}(p(\cdot \mid \doop(\boldsymbol{X}_{\mathcal{S}}=\mathbf{x}_{\mathcal{S}}^\star)),\widehat{p}_{t_0}(\cdot \mid \doop(\boldsymbol{X}_{\mathcal{S}}=\mathbf{x}_{\mathcal{S}}^\star)))\right] = \mathcal{T}(\mathbf{x}_{\mathcal{S}}^\star) \mathcal{O}(n^{-\frac{\beta}{q+2\beta}}\log^{\max(\frac{19}{2},\frac{\beta+3}{2})} n).
	\end{equation*}
	\endgroup
    	Moreover, we have
	\begin{equation*}
		\E_{\{\mathbf{X}_i\}_{i=1}^n} \left[\left\|\E_{p(\cdot \mid \doop(\boldsymbol{X}_{\mathcal{S}}=\mathbf{x}_{\mathcal{S}}^\star))}\left[\mathbf{x}\right] - \E_{\widehat{p}_{t_0}(\cdot \mid \doop(\boldsymbol{X}_{\mathcal{S}}=\mathbf{x}_{\mathcal{S}}^\star))}\left[\mathbf{x}\right]\right\|\right] = \mathcal{T}(\mathbf{x}_{\mathcal{S}}^\star) \mathcal{O}(n^{-\frac{\beta}{q+2\beta}}\log^{\max(11,\frac{\beta+6}{2})} n).
	\end{equation*}
\end{theorem}

Theorem~\ref{theorem:total_variation_bound_do_intervention} shows that the same local-dimension principle carries from the observational law to intervention targets, up to the intervention-specific shift factor $\mathcal{T}(\mathbf{x}_{\mathcal{S}}^\star)$. Thus the theorem controls total-variation recovery under intervention without claiming uniform performance over arbitrary intervention regimes.

Finally, we establish the validity of the resampling-based edge test.
\begin{theorem}
	\label{theorem:type1_control}
	Suppose the assumptions in Theorem \ref{theorem:total_variation_bound} hold and the distribution shift coefficient is uniformly bounded. We split the dataset $\{\mathbf{X}_{\cdot,i}\}_{i=1}^n$ into $\{\mathbf{X}_{i}\}_{i=1}^{n_1}$ for CEDM networks training and $\{\mathbf{X}_{i}\}_{i=n_1+1}^{n}$ for the inference in Algorithm \ref{alg:cedmi_single}. We denote $n_2 = n - n_1$. Under the null hypothesis, for any desired type~I error rate $\alpha \in [0,1]$, if $n_2 = o(n_1^{\frac{\beta}{1+q+2\beta}})$, we have
	$$\limsup_{n\to\infty}\mathbb{P}(p \leq \alpha) \leq \alpha.$$
\end{theorem}

Type~I error control hinges on exchangeability. We therefore adopt a sample-splitting scheme so that the fitted generative models are trained on data that are independent of the inference sample used to compute the observed and resampled statistics. This mirrors standard practice in resampling-based conditional independence testing, where an independent (or held-out) sample is typically required to fit the data-generating or conditional models under the null; see, for example, the CRT \citep{candes2018panning}, the CPT \citep{berrett2020conditional}, and the HRT \citep{tansey2022hrt}.

\section{Numerical examples}

\subsection{Distribution recovery}

We next assess whether encoding the true graph in diffusion generations improves recovery of interventional distributions. Throughout, we compare three approaches: (i) the proposed \emph{causality-encoded} diffusion model (CEDM); (ii) a general diffusion model that is agnostic to causal directionality, for which we use RePaint \citep{lugmayr2022repaint} as a generic conditioning device at sampling time; and (iii) VACA, a graph-based variational autoencoding approach designed for interventional and counterfactual queries \citep{sanchez2022vaca}.

\paragraph{Synthetic graphs.}
We consider $d=30$ variables partitioned into six 5-node slates,
$$\boldsymbol{Y}_1=(X_1,\ldots,X_5),\;\boldsymbol{Y}_2=(X_6,\ldots,X_{10}),\;\ldots,\;\boldsymbol{Y}_6=(X_{26},\ldots,X_{30}),$$
and study three slate-level DAGs:
\begin{enumerate}[(a)]
    \item \textit{Chain:} $\boldsymbol{Y}_1 \to \boldsymbol{Y}_2 \to \boldsymbol{Y}_3 \to \boldsymbol{Y}_4 \to \boldsymbol{Y}_5 \to \boldsymbol{Y}_6$.
    \item \textit{Hub:} $\boldsymbol{Y}_1$ is a common parent of all downstream slates, i.e., $\boldsymbol{Y}_1 \to \boldsymbol{Y}_k$ for $k=2,\ldots,6$.
    \item \textit{Random:} for each repetition, we sample a new DAG on $\{\boldsymbol{Y}_1,\ldots,\boldsymbol{Y}_6\}$, drawing each edge $\boldsymbol{Y}_i\to \boldsymbol{Y}_j$ ($i<j$) independently with probability $p=0.5$.
\end{enumerate}

For the \textit{Chain} and \textit{Hub} configurations, we draw $\boldsymbol{Y}_1$ from a zero-mean multivariate normal distribution with within-slate covariance $\Sigma_{ij}=0.5^{|i-j|}$. Each downstream slate is then generated via nonlinear structural equations with additive Gaussian noise that preserves the same within-slate correlation structure; noise terms are independent across slates. In the \textit{Random} configuration, each root slate is sampled from the same correlated Gaussian distribution, and each non-root slate is generated from all of its parent slates through nonlinear functions plus correlated Gaussian noise.

For each repetition, we train all methods using $n\in\{500,1000,2000,5000\}$ samples. For the \textit{Chain}, \textit{Hub}, and \textit{Random} settings, we intervene on $\boldsymbol{Y}_2$ and $\boldsymbol{Y}_5$ by fixing them at pre-specified values, and then generate samples for the remaining slates $\boldsymbol{Y}_1,\boldsymbol{Y}_3,\boldsymbol{Y}_4,\boldsymbol{Y}_6$. We assess recovery of the target interventional law by computing the squared maximum mean discrepancy (MMD) between the model-generated samples and an independent set of 5{,}000 reference samples drawn from the true interventional distribution.

\paragraph{\textit{Sachs} signalling network.}
We also consider the 11-node protein-signalling network of \citet{sachs2005causal} (Figure~\ref{fig:flow_cytometry_networks}(b)), generating data from nonlinear structural equations with additive Gaussian errors. Here we intervene on four proteins, namely PIP3, p38, Raf, and Erk, by fixing them at pre-specified values, and compare the generated samples for the remaining proteins with 5{,}000 reference samples from the corresponding true interventional distribution, again using squared MMD.

\begin{figure}[ht]
  \centering
  \includegraphics[width=\textwidth]{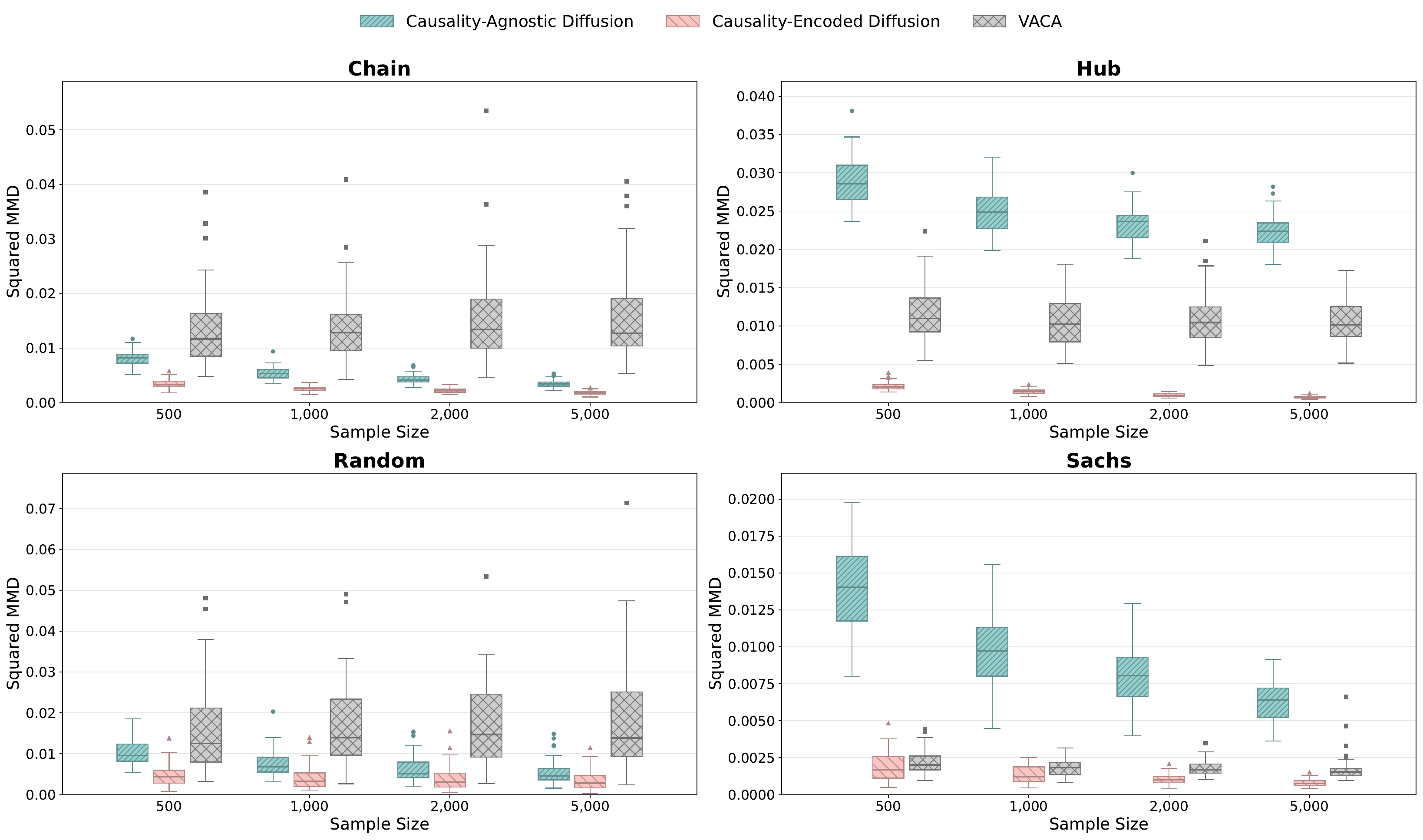}
  \caption{Simulation results on the \textit{Chain}, \textit{Hub}, \textit{Random}, and \textit{Sachs} structures. Boxplots compare squared MMD between the model-generated samples and 5{,}000 reference samples from the true interventional distribution across 50 independent repetitions for each training sample size. Lower values indicate better recovery of the target interventional law.}
  \label{fig:causal_structures_mmd_performance}
\end{figure}

\paragraph{Results.}
For each graph and each sample size $n \in \{500,1000,2000,5000\}$, we run 50 independent repetitions. Figure~\ref{fig:causal_structures_mmd_performance} summarises the resulting squared-MMD distributions. Across the four scenarios, CEDM typically attains the smallest MMD among the compared methods and therefore the best recovery of the target interventional law in this benchmark. The causality-agnostic diffusion baseline performs especially poorly in the \textit{Hub} and \textit{Sachs} settings, where the gap between the target $\doop$-distribution and the corresponding observational conditional law is larger, so generic conditioning at sampling time is least aligned with the causal question. VACA is expressly designed for DAG-based interventional and counterfactual tasks, but in these experiments it does not match the recovery accuracy of CEDM.

Some causal targets can also be summarised through conditional expectations, albeit less directly than through $\doop$-sampling. This motivates the additional experiment reported in Supplementary Section~1.1 \citep{chen2026supp}, where we study conditional expectation estimation directly. The same qualitative conclusion persists there: CEDM retains a systematic advantage even in a setting that is naturally suited to causality-agnostic conditional generation. Moreover, the gain is especially pronounced for sparse graphs, notably in the \textit{Chain} and \textit{Hub} settings, which is consistent with Theorem~\ref{theorem:total_variation_bound_do_intervention}: the recovery rate is governed by the maximum local dimension rather than the ambient dimension.

\subsection{Inference via CEDM}

We next study whether the proposed inference procedure is simultaneously well calibrated and powerful. To assess size, we consider a three-slate system,
$$\boldsymbol{Y}_1=(X_1,\ldots,X_{10}),\quad
\boldsymbol{Y}_2=(X_{11},\ldots,X_{20}),\quad
\boldsymbol{Y}_3=(X_{21},\ldots,X_{30}),$$
under two data-generating configurations, each corresponding to a distinct null hypothesis:
\begin{enumerate}[(a)]
    \item \textit{Chain:} the true slate-level DAG is $\boldsymbol{Y}_1 \to \boldsymbol{Y}_2 \to \boldsymbol{Y}_3$, and we test the absent edge $\boldsymbol{Y}_1 \to \boldsymbol{Y}_3$;
    \item \textit{Fork:} the true slate-level DAG is $\boldsymbol{Y}_1 \to \boldsymbol{Y}_2$ and $\boldsymbol{Y}_1 \to \boldsymbol{Y}_3$, and we test the absent edge $\boldsymbol{Y}_2 \to \boldsymbol{Y}_3$.
\end{enumerate}

\begin{figure}[H]
	\centering
	\includegraphics[width=\textwidth]{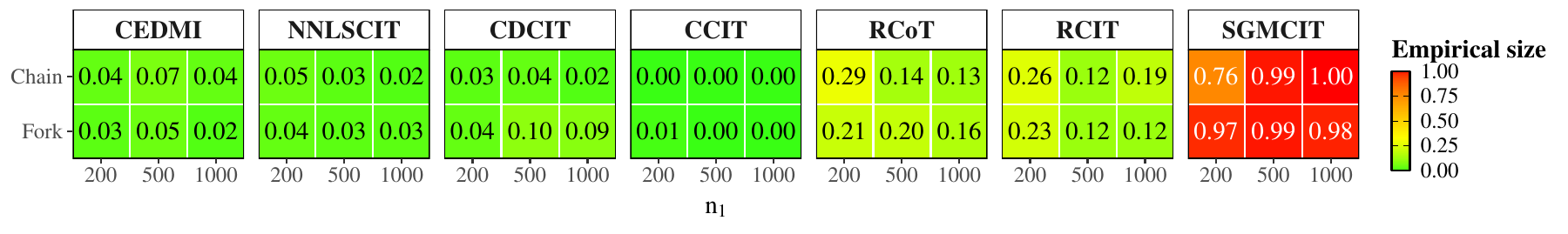}
	\caption{Empirical rejection rates (size) at nominal level 0.05 for CEDMI and competing conditional independence tests under the chain and fork null configurations.}
	\label{fig:inference_size_split}
\end{figure}

In both settings, variables are generated from nonlinear structural equations with Gaussian noise that is correlated within each slate and independent across slates. For each graph and each sample-size pair $(n_1,n_2)\in \{(200,200), (500,300), (1000,400)\}$, we generate 100 independent datasets. On each dataset, we train a CEDM that encodes the true causal structure, draw $N_{\textrm{mc}}=100$ conditional samples of $\boldsymbol{Y}_3$ given $(\boldsymbol{Y}_1,\boldsymbol{Y}_2)$ from the fitted model, and compute a $p$-value using Algorithm~\ref{alg:cedmi_single}. We estimate the empirical size at level $0.05$ by the proportion of replications for which $p \le 0.05$. The resulting rejection rates are reported in Figure~\ref{fig:inference_size_split}.

We benchmark CEDMI against established conditional independence tests spanning classifier-, kernel-, and generative-model--based approaches. CCIT~\citep{sen2017model} recasts the problem as classification with a nearest-neighbour bootstrap. NNLSCIT~\citep{li2023k} estimates conditional mutual information using local KNN resampling. RCIT and RCoT~\citep{strobl2019approximate} approximate KCIT~\citep{zhang2012kernel} via random Fourier features. CDCIT~\citep{yang2025conditional} combines the conditional randomisation test with conditional diffusion models, using a classifier-based conditional mutual information estimator as the test statistic. SGMCIT~\citep{ren2025score} uses score-based modelling with conditional sampling and an explicit goodness-of-fit check. For methods not requiring sample splitting, we pool the training and inference subsets.

As shown in Figure~\ref{fig:inference_size_split}, CEDMI remains close to the nominal $0.05$ level across both null configurations and all sample-size pairs. NNLSCIT is also well calibrated, while CDCIT is roughly calibrated overall but somewhat liberal in the fork setting. CCIT is conservative in these experiments. By contrast, RCoT, RCIT, and SGMCIT are clearly anti-conservative, often by a wide margin. Accordingly, the raw power numbers for markedly size-distorted procedures should be interpreted with caution. This calibration picture is important for interpreting power: comparisons are most meaningful among procedures that first achieve reasonable control of size. 

We next examine empirical power. We again use the three-slate system above, with edges from the first slate to the second present in all configurations. The third slate is generated according to the nonlinear structural equation
$$\boldsymbol{Y}_3
= \gamma\, \boldsymbol{f}(\boldsymbol{Y}_1)
+ \eta\, \boldsymbol{g}(\boldsymbol{Y}_2)
+ \gamma\eta\, \boldsymbol{h}(\boldsymbol{Y}_1,\boldsymbol{Y}_2) + \boldsymbol{\varepsilon},$$
where $\boldsymbol{f},\boldsymbol{g},\boldsymbol{h}$ are fixed nonlinear functions and the Gaussian noise is correlated within each slate and independent across slates. The parameters $\gamma$ and $\eta$ control the strengths of the direct effects from $\boldsymbol{Y}_1$ and $\boldsymbol{Y}_2$, respectively, to $\boldsymbol{Y}_3$.

\begin{figure}[ht]
	\centering
	\includegraphics[width=\textwidth]{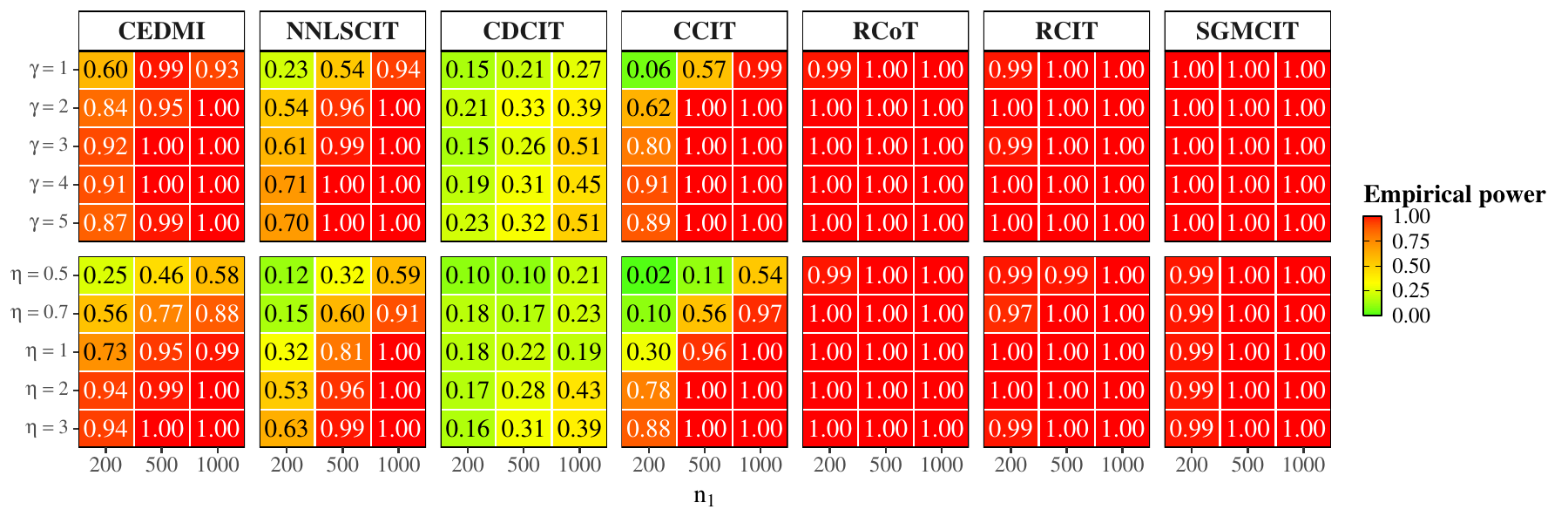}
	\caption{Empirical power at nominal level 0.05 for testing directed edges under varying signal strength. The top panel varies $\gamma$ when testing the null of no edge from $Y_1$ to $Y_3$; the bottom panel varies $\eta$ when testing the null of no edge from $Y_2$ to $Y_3$.}
	\label{fig:inference_power_split}
\end{figure}

For the top panel in Figure~\ref{fig:inference_power_split}, we consider alternatives to the null hypothesis of no directed edge from $\boldsymbol{Y}_1$ to $\boldsymbol{Y}_3$, varying $\gamma \in \{1.0, 2.0, 3.0, 4.0, 5.0\}$ with $\eta=1$. Larger values of $\gamma$ therefore correspond to stronger departures from the null. For the bottom panel in Figure~\ref{fig:inference_power_split}, we instead consider alternatives to the null of no directed edge from $\boldsymbol{Y}_2$ to $\boldsymbol{Y}_3$, varying $\eta \in \{0.5, 0.7, 1.0, 2.0, 3.0\}$ with $\gamma=1$. In each setting and for each sample-size pair, we generate 100 datasets, apply CEDMI and all competing tests, and estimate power as the proportion of replications with $p \le 0.05$.

Figure~\ref{fig:inference_power_split} reports the resulting empirical powers. For CEDMI, power increases with both sample size and signal strength $(\gamma$ or $\eta)$, reaching values close to one for moderate effects once $n_1 \ge 500$. NNLSCIT, CDCIT, and CCIT also become more powerful as the signal grows, but they typically trail CEDMI, especially at smaller sample sizes and weaker edge strengths. The near-unit rejection rates of RCoT, RCIT, and SGMCIT largely mirror their size distortions and therefore should not be read as genuine evidence of superior sensitivity. Taken together, the size and power experiments indicate that CEDMI attains strong power against meaningful alternatives while preserving the calibration needed for reliable inference.

For completeness, the Supplementary Materials \citep{chen2026supp} also report a no-sample-splitting variant in which the same dataset is used for both CEDM fitting and CEDMI. In these experiments, type~I error remains well controlled empirically and power is typically further improved.

\subsection{Flow cytometry data study}

\begin{figure}[htbp]
	\centering
	\begin{subfigure}{0.32\textwidth}
    \centering
    \resizebox{\linewidth}{!}{
		\begin{tikzpicture}[
  roundnode/.style={circle, draw=black, fill=gray!20, thick, minimum size=10mm, inner sep=1pt},
  arrow/.style={-Latex, thick, draw=black!65},
  bluearrow/.style={-Latex, thick, blue, dashed},
  every node/.style={font=\small}
]

% Circle radius
\def\R{3.2}

% Nodes on a circle (angles in degrees; adjust order if you want a different clockwise listing)
\node[roundnode] (PIP3) at ( 100:\R) {PIP3};
\node[roundnode] (PLCg) at ( 60:\R) {PLCg};
\node[roundnode] (PIP2) at ( 20:\R) {PIP2};
\node[roundnode] (Raf)  at (-23:\R) {Raf};
\node[roundnode] (Mek)  at (-54:\R) {Mek};
\node[roundnode] (Erk)  at (-90:\R) {Erk};
\node[roundnode] (Akt)  at (-126:\R) {Akt};
\node[roundnode] (PKA)  at (-162:\R) {PKA};
\node[roundnode] (PKC)  at ( 162:\R) {PKC};
\node[roundnode] (p38)  at ( 120:\R) {p38};
\node[roundnode] (JNK)  at (  0:\R) {JNK}; 

% Edges (grey)
\draw[arrow] (PIP3) -- (PIP2);
\draw[arrow] (PLCg) -- (PIP2);

\draw[arrow] (Mek) -- (Erk);
\draw[arrow] (PKA) -- (Erk);
\draw[arrow] (PKA) -- (Akt);

\draw[arrow] (PKA) -- (Mek);
\draw[arrow] (PKA) -- (Raf);
\draw[arrow] (PKA) -- (JNK);
\draw[arrow] (PKA) -- (p38);

\draw[arrow] (PKC) -- (Mek);
\draw[arrow] (PKC) -- (Raf);
\draw[arrow] (PKC) -- (JNK);
\draw[arrow] (PKC) -- (p38);

\draw[arrow] (Raf) -- (Mek);

% Blue dashed edges
\draw[bluearrow] (PIP3) -- (PLCg);
\draw[bluearrow] (PIP3)  -- (Akt);
\draw[bluearrow] (PIP2)  -- (PKC);
\draw[bluearrow] (PLCg)  -- (PKC);

\end{tikzpicture}
}
		\caption{Consensus}
	\end{subfigure}
	\hfill
	\begin{subfigure}{0.32\textwidth}
		\centering
        \resizebox{\linewidth}{!}{
		\begin{tikzpicture}[
  roundnode/.style={circle, draw=black, fill=gray!20, thick, minimum size=10mm, inner sep=1pt},
  arrow/.style={-Latex, thick, draw=black!65},
  bluearrow/.style={-Latex, thick, blue, dashed},
  every node/.style={font=\small}
]

% Circle radius
\def\R{3.2}

% Nodes on a circle (angles in degrees; adjust order if you want a different clockwise listing)
\node[roundnode] (PIP3) at ( 100:\R) {PIP3};
\node[roundnode] (PLCg) at ( 60:\R) {PLCg};
\node[roundnode] (PIP2) at ( 20:\R) {PIP2};
\node[roundnode] (Raf)  at (-23:\R) {Raf};
\node[roundnode] (Mek)  at (-54:\R) {Mek};
\node[roundnode] (Erk)  at (-90:\R) {Erk};
\node[roundnode] (Akt)  at (-126:\R) {Akt};
\node[roundnode] (PKA)  at (-162:\R) {PKA};
\node[roundnode] (PKC)  at ( 162:\R) {PKC};
\node[roundnode] (p38)  at ( 120:\R) {p38};
\node[roundnode] (JNK)  at (  0:\R) {JNK}; 

% Edges (grey)
\draw[arrow] (PIP3) -- (PIP2);
\draw[arrow] (PLCg) -- (PIP2);

\draw[arrow] (Mek) -- (Erk);
\draw[arrow] (PKA) -- (Erk);
\draw[arrow] (PKA) -- (Akt);

\draw[arrow] (PKA) -- (Mek);
\draw[arrow] (PKA) -- (Raf);
\draw[arrow] (PKA) -- (JNK);
\draw[arrow] (PKA) -- (p38);

\draw[arrow] (PKC) -- (Mek);
\draw[arrow] (PKC) -- (Raf);
\draw[arrow] (PKC) -- (JNK);
\draw[arrow] (PKC) -- (p38);

\draw[arrow] (Raf) -- (Mek);

% Blue dashed edges
\draw[bluearrow] (PLCg) -- (PIP3);
\draw[bluearrow] (Erk)  -- (Akt);
\draw[bluearrow] (PKC)  -- (PKA);

\end{tikzpicture}
}
		\caption{Sachs \textit{et al.}}
	\end{subfigure}
	\hfill
	\begin{subfigure}{0.32\textwidth}
		\centering
        \resizebox{\linewidth}{!}{
		\begin{tikzpicture}[
  roundnode/.style={circle, draw=black, fill=gray!20, thick, minimum size=10mm, inner sep=1pt},
  arrow/.style={-Latex, thick, draw=black!65},
  bluearrow/.style={-Latex, thick, blue, dashed},
  every node/.style={font=\small}
]

% Circle radius
\def\R{3.2}

% Nodes on a circle (angles in degrees; adjust order if you want a different clockwise listing)
\node[roundnode] (PIP3) at ( 100:\R) {PIP3};
\node[roundnode] (PLCg) at ( 60:\R) {PLCg};
\node[roundnode] (PIP2) at ( 20:\R) {PIP2};
\node[roundnode] (Raf)  at (-23:\R) {Raf};
\node[roundnode] (Mek)  at (-54:\R) {Mek};
\node[roundnode] (Erk)  at (-90:\R) {Erk};
\node[roundnode] (Akt)  at (-126:\R) {Akt};
\node[roundnode] (PKA)  at (-162:\R) {PKA};
\node[roundnode] (PKC)  at ( 162:\R) {PKC};
\node[roundnode] (p38)  at ( 120:\R) {p38};
\node[roundnode] (JNK)  at (  0:\R) {JNK}; 

% Edges (grey)
\draw[arrow] (PIP3) -- (PIP2);
\draw[arrow] (PLCg) -- (PIP2);

\draw[arrow] (Mek) -- (Erk);
\draw[arrow] (PKA) -- (Erk);
\draw[arrow] (PKA) -- (Akt);

\draw[arrow] (PKA) -- (Mek);
\draw[arrow] (PKA) -- (Raf);
\draw[arrow] (PKA) -- (JNK);
\draw[arrow] (PKA) -- (p38);

\draw[arrow] (PKC) -- (Mek);
\draw[arrow] (PKC) -- (Raf);
\draw[arrow] (PKC) -- (JNK);
\draw[arrow] (PKC) -- (p38);

\draw[arrow] (Raf) -- (Mek);

\draw[arrow] (PIP3) -- (PLCg);
\draw[arrow] (PIP3)  -- (Akt);
\draw[arrow] (PIP2)  -- (PKC);
\draw[arrow] (PLCg)  -- (PKC);

\draw[arrow] (Erk)  -- (Akt);
\draw[arrow] (PKC)  -- (PKA);

\end{tikzpicture}
}
		\caption{Super-DAG}
	\end{subfigure}
	\caption{(a) Consensus network, according to \citep{sachs2005causal}; (b) Reconstructed signalling network by \citep{sachs2005causal}; (c) Super-DAG obtained by taking the union of (a) and (b), with the edge between PIP3 and PLCg oriented as PIP3 $\to$ PLCg. In panels (a) and (b), blue dashed arrows indicate edges on which the two networks disagree.}
	\label{fig:flow_cytometry_networks}
\end{figure}
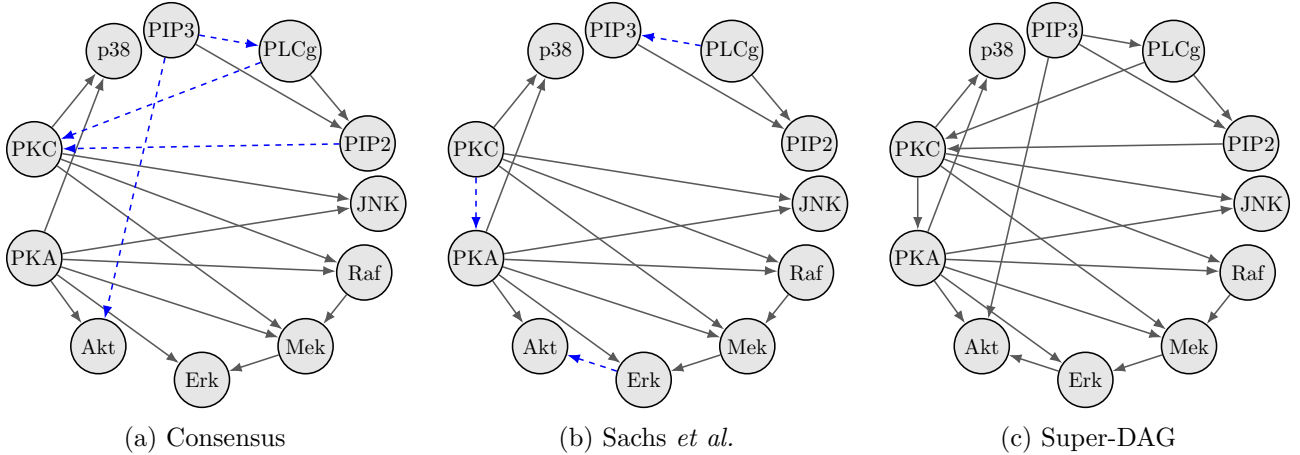

In this subsection, we apply CEDMI to the protein--signalling flow cytometry dataset of \citet{sachs2005causal}. In that study, the authors first assembled a literature-based ``consensus'' signalling network over 11 phosphorylated proteins (Figure~\ref{fig:flow_cytometry_networks}(a)), and then inferred a data-driven Bayesian network from the same single-cell measurements via Bayesian network structure learning (Figure~\ref{fig:flow_cytometry_networks}(b)). The two graphs coincide on most edges but disagree on a number of connections. Our aim is to use CEDMI to provide formal statistical evidence for or against these disputed edges using the flow cytometry data.

\paragraph*{Dataset.}
The dataset comprises single-cell measurements from human primary CD4$^+$ T cells. Using multicolour flow cytometry, the experiment records the activity (phosphorylation state or abundance) of 11 signalling molecules: Raf, Mek, PLCg, PIP2, PIP3, Erk, Akt, PKA, PKC, p38, and JNK \citep{sachs2005causal}. Each observation corresponds to a single cell, for which the measured levels of all 11 proteins are available. Following \citet{mooij2013cyclic}, we use a continuous, pre-processed version of the data, consisting of $n=1755$ single-cell measurements on the same variables. This dataset, together with the associated consensus network and the Bayesian network reported by \citet{sachs2005causal}, has become a standard benchmark for causal network inference.

Figure~\ref{fig:flow_cytometry_networks}(c) displays the super-DAG that we adopt as the background graph for inference. It contains every directed edge appearing in either the consensus network or the network learned by Sachs et al., except for the pair (PIP3, PLCg), for which we fix the orientation as PIP3 $\to$ PLCg. Constructing a super-DAG and subsequently testing individual edges is a common strategy in the DAG-inference literature \citep{li2020likelihood,chen2024discovery}. In our analysis, we test only those edges on which panels (a) and (b) disagree, excluding PIP3 $\to$ PLCg; consequently, the arbitrary orientation chosen for the PIP3--PLCg pair has no bearing on the validity of the remaining tests.

\paragraph*{Results.}
Let $\mathcal{E}_0$ denote the edge set of the super-DAG in Figure~\ref{fig:flow_cytometry_networks}(c), and let $\mathcal{E}_1 \subseteq \mathcal{E}_0$ denote the subset of disputed edges. For each $e \in \mathcal{E}_1$, we apply CEDMI using the super-DAG as background knowledge. Specifically, for each disputed linkage we form the corresponding null graph by deleting that linkage from $\mathcal{E}_0$ (and, for $(\text{PIP2}, \text{PLCg}) \to \text{PKC}$, deleting both incoming PKC edges jointly), then refit the CEDM under that null graph and compare the observed test statistic with its Monte Carlo null distribution based on $N_{\mathrm{mc}}=500$ synthetic draws. The resulting null distributions and observed statistics are reported in the Supplementary Materials \citep{chen2026supp}.

We also apply NNLSCIT, CDCIT, and CCIT, selected on the basis of their performance in the simulation study. For the linkages $(\text{PIP2}, \text{PLCg}) \to \text{PKC}$ and $\text{PKC} \to \text{PKA}$, note that PKC and PKA have no parents other than the variables directly involved in the corresponding tests. Consequently, under the null hypothesis, valid reference distributions can also be obtained by independently permuting the observations of PKC and PKA, respectively. The resulting permuted statistics yield empirical null distributions and corresponding permutation $p$-values.

The final testing outcomes are summarised in Table~\ref{tab:summary_cytometry}. A tick indicates that an edge appears in the corresponding graph or that the corresponding edge-deletion null is rejected by the method, whereas a cross indicates that it is absent from the reference graph or that the method does not reject the null.

\begin{table}[ht]
	\centering
	\small
	\begin{tabular}{lccccccc}
		\toprule
		Edges                         & Consensus & Sachs  & Permutation        & CEDMI & NNLSCIT & CDCIT & CCIT \\
		\midrule
		(PIP2,PLCg) $\rightarrow$ PKC & $\checkmark$      & $\times$ & $\times$ & $\times$  & $\times$ & $\times$ & $\times$      \\
		PKC $\rightarrow$ PKA         & $\times$     & $\checkmark$  & $\times$ & $\times$    & $\times$ & $\times$ & $\times$     \\
		PIP3 $\rightarrow$ Akt        & $\checkmark$      & $\times$ &   \textemdash & $\times$  & $\times$ & $\times$ & $\times$ \\
		Erk $\rightarrow$ Akt         & $\times$     & $\checkmark$   & \textemdash & $\checkmark$ & $\checkmark$ & $\checkmark$ & $\checkmark$ \\
		\bottomrule                         
	\end{tabular}
	\caption{Summary of edge inclusion decisions for the four disputed linkages in the flow cytometry signalling network.}
	\label{tab:summary_cytometry}
\end{table}

\begin{figure}[ht]
	\centering
	\includegraphics[width=\textwidth]{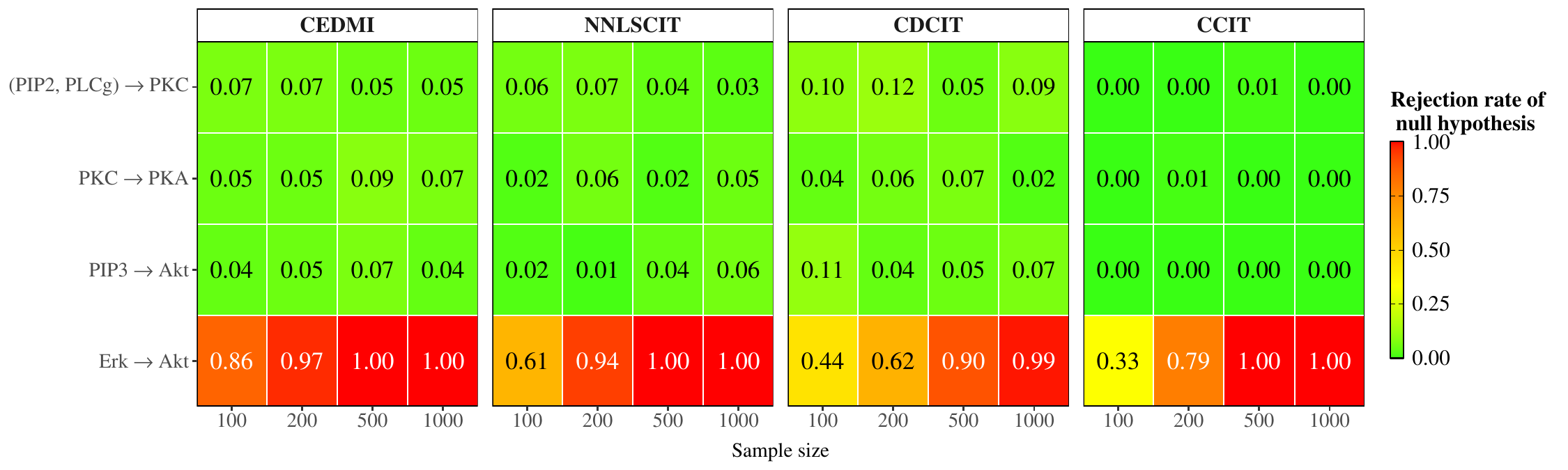}
	\caption{Rejection rates for tests of the four disputed linkages in the flow cytometry network across four sample sizes and four competing methods. Columns correspond to CEDMI, NNLSCIT, CDCIT, and CCIT, and rows correspond to the linkages $(\text{PIP2},\text{PLCg}) \to \text{PKC}$, $\text{PKC} \to \text{PKA}$, $\text{PIP3} \to \text{Akt}$, and $\text{Erk} \to \text{Akt}$. Within each panel, the colour and printed value give the empirical rejection rate over 100 repeated subsamples at sample sizes $100$, $200$, $500$, and $1000$.}
	\label{fig:edge_test_heatmap}
\end{figure}

To assess stability across sample sizes, we repeatedly drew subsamples from the full dataset, considering sample sizes of 100, 200, 500, and 1000, with 100 repetitions for each size. We then applied CEDMI, NNLSCIT, CDCIT, and CCIT to the disputed linkages. Figure~\ref{fig:edge_test_heatmap} summarises the resulting rejection rates for the null hypothesis that the corresponding linkage is absent.

Our real-data analysis yields three main conclusions. First, for three of the four disputed linkages, CEDMI and the three comparison methods do not support the corresponding consensus-network edges. Notably, for the two linkages for which a direct permutation test is available, all four methods agree with the permutation-based conclusion. Since the literature-based consensus network of \citet{sachs2005causal} is often treated as a working ground truth for this dataset \citep{sen2017model,li2023k,yang2025conditional}, these findings suggest that caution is warranted when interpreting results against that network as though it were definitive.

Second, across repeated subsamples, CEDMI maintains low rejection rates for the linkages whose edge-deletion nulls are not rejected, while producing the strongest and most stable evidence against the null that omits $\text{Erk} \to \text{Akt}$. This mirrors the calibration-and-power pattern seen in the simulations.

Third, a notable finding is the support for the edge $\text{Erk} \to \text{Akt}$. Although this edge does not appear in the canonical T-cell signalling diagram, it is consistent with experimental reports of ERK--AKT cross-talk in mammalian cells, including ERK-dependent activation of AKT in liver cancer cells and other systems \citep{kim2020malignancy,khundmiri2007ouabain,stulpinas2023crosstalk}. Accordingly, our findings point to a biologically plausible Erk--Akt interaction in this T-cell signalling context, although direct experimental validation in naive CD4$^+$ T cells would still be required.

\section{Discussion}
This paper studies diffusion learning in the presence of graph information. The main point is not merely that a known DAG can be encoded into the architecture, but that doing so changes both what the model can represent and how hard the statistical problem is. By replacing a single joint score with nodewise conditional scores, CEDM yields a generative procedure with interventional semantics and recovery guarantees driven by the maximum local parent--child dimension rather than the ambient dimension. The companion procedure CEDMI then turns that generator into a null-sampling device for targeted edge tests. A limitation is the reliance on a known graph. Thus a natural next step is a unified diffusion-based approach to causal discovery that couples structure learning with causality-encoded score estimation and uncertainty quantification. We provide additional discussions on the performance of CEDMI when the working graph is misspecified under the null hypothesis in the Supplementary Materials \citep{chen2026supp}.

\section*{Acknowledgements}
This work was supported in part by NSF grant DMS-2513668 and NIH grants R01AG069895, R01AG065636, R01AG074858, and U01AG073079.

\bibliographystyle{unsrtnat} 
\bibliography{reference}

\end{document}